\documentclass[12pt,preprint]{aastex}

\shorttitle{fuzzy dark matter}
\shortauthors{Burkert}

\begin{document}

\title{Fuzzy Dark Matter and Dark Matter Halo Cores}

\author{A. Burkert\altaffilmark{1,2}}

\altaffiltext{1}{University Observatory Munich (USM), Scheinerstrasse 1, 81679 Munich,
Germany}
\altaffiltext{2}{Max-Planck-Institut f\"ur extraterrestrische Physik (MPE), Giessenbachstr. 1, 85748 Garching, Germany}

\email{burkert@usm.lmu.de}

\newcommand\msun{\rm M_{\odot}}
\newcommand\lsun{\rm L_{\odot}}
\newcommand\msunyr{\rm M_{\odot}\,yr^{-1}}
\newcommand\be{\begin{equation}}
\newcommand\en{\end{equation}}
\newcommand\cm{\rm cm}
\newcommand\kms{\rm{\, km \, s^{-1}}}
\newcommand\K{\rm K}
\newcommand\etal{{\rm et al}.\ }
\newcommand\sd{\partial}

\begin{abstract}
Whereas cold dark matter (CDM) simulations predict central dark matter cusps with densities that diverge as
$\rho$(r)$\sim$ 1/r observations often indicate constant density cores with finite central densities $\rho_0$ and
a flat density distribution within a core radius r$_0$. This paper investigates whether this core-cusp problem
can be solved by fuzzy dark matter (FDM), a hypothetical particle with a mass of order m$\approx$10$^{-22}$eV and 
a corresponding de Broglie wavelength on astrophysical scales. We show that galaxies with CDM halo virial masses 
M$_{vir} \leq 10^{11}$M$_{\odot}$ follow two core scaling relations. In addition to the well known universal 
core column density $\Sigma_0 \equiv \rho_0 \times$r$_0$ = 75 M$_{\odot}$pc$^{-2}$ core
radii increase with virial masses as r$_0 \sim$ M$_{vir}^{\gamma}$ with $\gamma$ of order unity. 
Using the simulations by Schive et al. (2014) we demonstrate
that FDM can explain the r$_0$-M$_{vir}$ scaling relation if the virial masses of the observed galaxy
sample scale with formation redshift z as M$_{vir}\sim$(1+z)$^{-0.4}$. The observed constant $\Sigma_0$
is however in complete disagreement with FDM cores which are characterised by a steep dependence
$\Sigma_0 \sim$r$_0^{-3}$, independent of z. 
More high-resolution simulations are now required to confirm the simulations of Schive et al. (2014) and explore
especially the transition region between the soliton core and the surrounding halo.
If these results hold,
FDM can be ruled out as the origin of observed dark matter cores and other physical processes 
are required to account for their formation.  

\end{abstract}


\section{Introduction}
One of the strongest constraints for the $\Lambda$ cold-dark-matter (CDM) model of cosmic structure formation
(Blumenthal et al. 1984, White \& Frenk 1991) is the empirical result of numerous numerical simulations 
that CDM halos have universal density distributions that are well fit by an NFW profile 
(Navarro, Frenk \& White 1996; for a review see Kuhlen, Vogelsberger \& Angulo 2012) 

\begin{equation}
\rho_{NFW} (r)= \rho_s \left(\frac{4 r_s^3}{r (r+r_s)^2} \right)
\end{equation}

\noindent with r$_s$ the dark halo scale radius and $\rho_s$ the density at r=r$_s$. For r$\ll$r$_s$ the density diverges as 
$\rho\sim $r$^{-1}$
which is called the central cusp.  Hydrostatic equilibrium requires that for an isotropic velocity distribution
the dark matter particle velocity dispersion $\sigma$ within this cusp
decreases towards the center as $\sigma \sim  r^{1/2}$. Dark matter cusps are therefore kinematically cold which can
be understood as a relict of the first structures that formed in the early, dense Universe when cold dark matter particles by definition
still had small random velocities.
In contrast to this fundamental property of CDM halos, observations, especially of lower-mass, dark matter dominated galaxies, often show 
rotation curves with a shape at small radii that points to flat inner dark matter density
distributions. These so called dark matter cores are characterised by a finite central density $\rho_0$ and a
flat density profile within a core radius r$_0$, 
reminiscent of self-gravitating isothermal spheres (e.g. Moore, 1994; Burkert 1995; 
Gentile et al. 2004; de Blok 2010; Pontzen \& Governato 2014; Fern\'{a}ndez-Hern\'{a}ndez et al. 2019; Genzel et al. 2020; for reviews with
references see Weinberg et al. 2015, Li et al. 2020 and di Paolo \& Salucci 2020). Various empirical density distributions have been proposed to
fit these cores. A profile that is frequently used is (Burkert 1995)

\begin{equation}
\rho_B (r) = \rho_0 \times \frac{r_0^3}{(r+r_0)(r^2+r_0^2)}
\end{equation}

\noindent Within r$_0$ the Burkert profile follows an isothermal
sphere. For larger radii it transits into the typical NFW profile with its characteristic r$^{-3}$ decline.

The origin of the CDM core-cusp problem is highly debated.  One class of models invokes violent fluctuations
of the gravitational potential in the inner regions of galaxies, caused e.g. by  baryonic processes like perturbations due to
the clumpy, turbulent interstellar medium or
strong galactic winds that remove a large fraction of an early gravitationally dominent gas component
(Navarro, Eke \& Frenk 1996; Governato et al. 2012; Teyssier et al. 2013; di Cintio et al. 2014; Ogiya \& Mori 2014; Pontzen \& Governato 2014; Chan et al. 2015; El-Zant et al. 2016; 
Peirani et al. 2017; Ben\'{i}tez-Llambay et al. 2019; Freundlich et al. 2020). 
Another even more fascinating possibility is however that the core-cusp problem points towards
hidden properties of the dark matter particle itself that are not taken into account in standard cosmological simulations.

One such scenario that has received much attention recently is fuzzy dark matter (FDM; Hu et al. 2000). The FDM model assumes that
dark matter particles are axions with a mass of order m$\approx$10$^{-22}$eV and 
a corresponding de Broglie wavelength as large as the typical galactic scale length
(Goodman 2000; Schive et al. 2014a,b; Hui et al. 2017, Bernal et al. 2018). The observed dark matter cores would then be
soliton cores, resulting from a balance between quantum pressure due to the uncertainty principle and gravity and the observed core
properties would directly trace the FDM particle mass m.
Given m, the core properties are completely determined by solving the coupled Schr\"odinger-Poisson equation (Widrow \& Kaiser 1993).

Several groups have compared the predicted FDM core structure with observations.
Deng et al. (2018), for example, examined a large class of theoretical light scalar DM models, 
governed by some potential V and assuming a scalar that is complex with a global $U$(1) symmetry. They demonstrated that within the framework of 
their analytical model there does not exist one single axion mass that can explain the observed
large range of core radii r$_0$ and at the same time reproduce the
observed core scaling relation $\rho_0 \sim r_0^{-1}$ (Burkert 2015; Kormendy \& Freeman 2016; Rodrigues 
et al. 2017). This interesting result however does not take into account the build-up of dark halos 
by cosmic structure formation.
The first self-consistent cosmological 3D simulation of FDM halo formation was presented by
Schive et al. (2014a,b). They confirmed that all halos develop a distinct, gravitationally self-bound solitonic core with
a universal core density distribution. For radii r $\leq 3 \times$r$_0$ it can be well fitted by the empirical relation
\begin{equation}
\rho (r) = 0.019 \times \left( \frac{m}{10^{-22}{\rm eV}} \right)^{-2} \times \left( \frac{(r_0/{\rm kpc})^{-1}}{(1+9.1 \times 10^{-2} (r/r_0)^2)^2} \right)^{4} \frac{{\rm M}_{\odot}}{{\rm pc}^3}.
\end{equation}

\noindent In addition, the core radius r$_0$ scales with the halo virial mass and cosmological redshift z as

\begin{equation}
r_0 = 1.6 \times q(z) \times(1+z)^{-0.5} \left( \frac{m}{10^{-22}{\rm eV}} \right)^{-1} \left( \frac{M_{vir}}{10^9 {\rm M}_{\odot}} \right)^{-1/3}{\rm kpc}
\end{equation}

\noindent with q(z)=[$\zeta$(z)/$\zeta$(0)]$^{1/6}$ and
$\zeta$(z) = $\left( 18 \pi^2 + 82(\Omega_m(z)-1)-39(\Omega_m(z)-1)^2 \right)/\Omega_m(z)$. 
For a present-day matter density parameter 
$\Omega_m$(z=0) = 0.315  (Planck Collaboration 2018) the value of q decreases from 1 for z=0 
to 0.9 at very high redshifts z$\geq$100. This change is much smaller than the observational uncertainties in 
determining halo core properties. We therefore will adopt q(z)=1 throughout this paper.
For r$>$3$\times$r$_0$ the density distribution approaches the characteristic r$^{-3}$ NFW density decline (equation 1).

The results of Schive et al. (2014) have been confirmed lateron by cosmological zoom-in simulations of Veltmaat et al. (2018). 
Interestingly, these authors also detected strong largely undamped quasi-normal oscillations within the soliton core. Outside the
core and within the halo virial radius de Broglie fluctuations generate a granular structure with order of unity
density fluctuations, resulting from wave interference.

The Schive et al. (2014a,b) soliton core profile (equation 3) has been used 
to derive limits on the FDM particle mass m (e.g. Hui et al. 2017; Marsh \& Pob 2015;
Gonz\'{a}lez-Morales et al. 2017, Schutz 2020). Calabrese \& Spergel (2016) analysed two of the faintest, strongly dark matter
dominated Milky Way dwarf galaxies, Draco II and Triangulum II, and found
m = 3.7 - 5.6 $\times 10^{-22}$eV. This result however relies on the assumption that the stellar component is 
completely embedded within the soliton core and that these diffuse satellite galaxies with galactocentric distances of
20 kpc (Laevens et al. 2015a) and 26 kpc (Laevens et al. 2015b), respectively, are not strongly tidally 
perturbed and in virial equilibrium. Safarzadeh \& Spergel (2020) presented an analyses of two more distant 
Milky Way dwarf spheroidals, 
Sculptor and Fornax, with galactocentric distances of  88 kpc and and 138 kpc, respectively (Kormendy \& Freeman 2016). They inferred
axion masses of order $10^{-21}$ eV with some dependence on the unknown halo virial mass.
Wasserman et al. (2019) looked at the ultra-diffuse, strongly dark matter dominated galaxy Dragonfly 44 and derived soliton masses of order 
$\sim 3 \times 10^{-22}$eV. Li et al. (2020) demonstrated that a soliton core, corresponding to a
boson mass of $\sim$2-7$\times$10$^{-22}$eV would help to explain the origin of the Milky Way central molecular zone dynamics, which requires
a dense, compact central mass concentration. 
Finally,  Davies \& Mocz (2020) showed that supermassive black holes in the centers of galaxies 
could affect the density profile of soliton cores, constraining the FDM particle mass.  
In summary, these studies indicate a FDM particle mass in the 
range of $10^{-22}$eV $\leq$m$\leq$ 10$^{-21}$eV. 

Other groups explored the visible effects which the of order unity FDM density fluctuations (Veltmaat et al. 2018) would have 
on structures within galaxies. Marsh \& Niemeyer (2019) showed that these fluctuations could heat and destroy star clusters.
They applied their model to an old star cluster, detected in the core of the ultrafaint dwarf
galaxy Eridanus II and found a lower limit of
m$\approx$10$^{-21}$eV, required in order for the cluster to have survived up to now.
Amorisco \& Loeb (2018) demonstrated that FDM density fluctuations
could thicken thin stellar streams, an effect that
could be detected with GAIA (see also Church et al. 2019). Bar-Or et al. (2019) and El-Zant et al. (2020)
lateron showed that the
stochastic FDM density fluctuations  can scatter stars and black holes, resulting in a diffusion through
phase space that should affect dynamical friction and the inspiraling of supermassive black holes and globular clusters in galaxies.

So far, the FDM core structure as predicted by the Schive et al. (2014a,b) simulations has been compared mainly with the structure of a few
Milky Way satellite galaxies. Here we go the next step and investigate whether it can explain
consistently the core scaling relations of a large sample of galaxies. We start by demonstrating that there are actually two core scaling relations
that need to be explained by FDM  or any other scenario of core formation. We then add a new element that has been
neglected so far,
the cosmic redshift dependence of soliton core properties. This is crucial!
As shown by equation 4 the radius of a soliton core depends not only on the axion mass m and the virial mass M$_{vir}$
but also on the redshift z when the dark halo core formed, i.e. when the halo stopped growing through accretion
by decoupling from the cosmic web. This is especially relevant for dwarf spheroidal
satellite galaxies that have preferentially been studied so far, but also for cluster galaxies in general. 
In section 2 we discuss the two core scaling relations and show that FDM cores predict completely different correlations than observed
for halos that could grow till z=0.
Section 3 then demonstrates that adding z as a second free parameter in addition to m can explain the origin
of equation 4. It however cannot solve the problem with first core scaling relation (equation 3).
Section 4 summarizes the results.

\section{Comparing observed core scaling relations with FDM predictions for z=0}

Cores like those given by equation 2 are characterized by two parameters, the core radius r$_0$ and the central density $\rho_0$.
In principle, both parameters could vary independently from galaxy to galaxy, depending for example on the details of the core formation history. 
The situation is however more interesting.

Dark halo core properties have been determined preferentially for lower-mass galaxies 
with stellar masses M$_* \leq 10^{10}$ M$_{\odot}$ which often have small baryon fractions.
These observations have revealed a tight correlation between r$_0$ and  $\rho_0$ that provides important constraints for any theoretical model of core formation.
As shown by the red points and the solid and dashed lines in the left panel of figure 1, 
all observations are consistent with a constant so called
core column density (Salucci \& Burkert 2000; Donato et al. 2009; Burkert 2015; Kormendy \& Freeman 2016)

\begin{equation}
\Sigma_{DM} \equiv \rho_0 \times r_0 = 75^{+55}_{-45} {\rm M}_{\odot} {\rm pc}^{-2}
\end{equation}

\noindent Note that the term column density is misleading as the product $\rho_0 \times r_0$ differs
from the integral $\int_{-R_{vir}}^{R_{vir}} \rho(r) dr$.

\begin{figure}[!ht]
\epsscale{1.}
\plotone{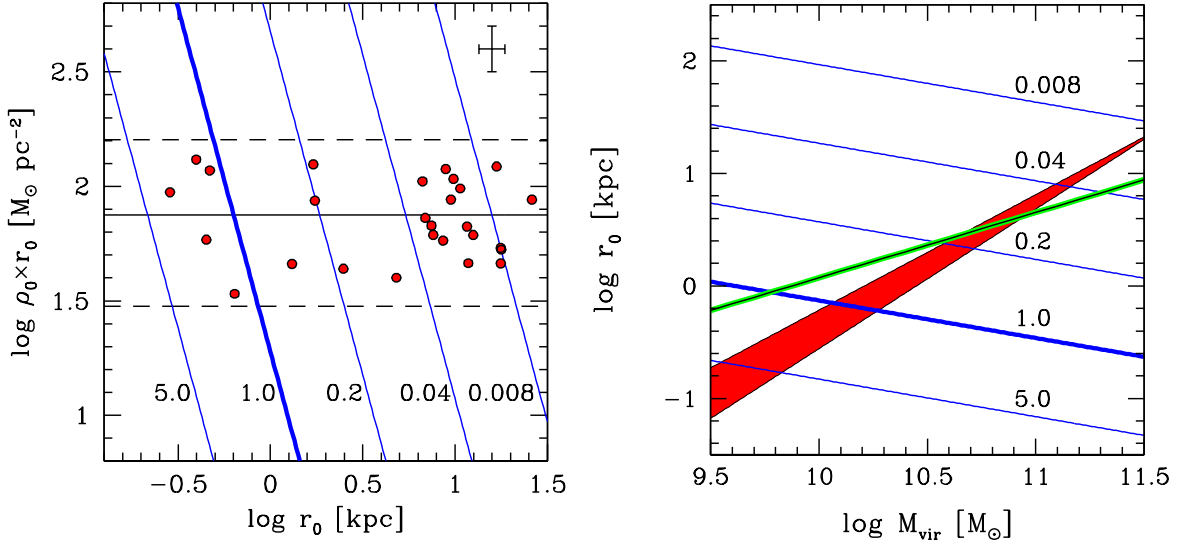}
\caption{
 \label{fig1}
{Observationally inferred dark halo core properties are compared with the predictions of the fuzzy dark matter
model. The red points in the left panel show the observed core column densities versus the corresponding core radii. The black solid
horizontal line depicts the average value of 75 M$_{\odot}$ pc$^{-2}$ and the horizontal dashed lines show the observed scatter.
The errorbar in the upper right corner shows the typical uncertainties.
Blue lines denote the predictions for soliton cores. The FDM cores follow a completely different trend compared
to the observations.  The numbers associated
with each blue line in the left and right panel depict the corresponding particle mass m in units of 
the standard value of $10^{-22}$eV. The thick solid blue line corresponds to this standard value.
The right panel shows the correlation between core radius and dark halo virial mass. The red area depicts the observationally inferred
correlation, adopting a stellar-to-dark matter mass conversion as predicted from abundance matching, combined with a constant stellar mass-to-light ratio
and assuming a halo formation time in the range z=0 (lower boundary) to z=8 (upper boundary).
Galaxies in this redshift range should populate this area. The green line shows the prediction from a universal rotation curve analyses of disk
galaxies by Salucci et al. (2007). The redshift zero FDM predictions (blue lines) follow again a completely different trend,
with r$_0$ decreasing with increasing M$_{vir}$.}}
\end{figure}

In addition to equation 5, Kormendy \& Freeman (2016) found that core radii increase with stellar mass M$_*$ as

\begin{equation}
r_0 = 5 {\rm kpc} \left( \frac{M_*}{L_B} \right)_{\odot}^{-0.446} \left( \frac{M_*}{10^9{\rm M}_{\odot}} \right)^{0.446}
\end{equation}

\noindent where L$_B$ is the blue luminosity and $(M_*/L_B)_{\odot}$ is the stellar mass to blue luminosity ratio in solar units.
For masses M$_* \leq 10^{10}$M$_{\odot}$ the star-to-dark matter conversion factor, predicted by abundance 
matching (Moster et al. 2013, 2018) is

\begin{equation}
\left( \frac{M_*}{M_{vir}} \right) = 2 \epsilon_N \left( \frac{M_{vir}}{M_1} \right)^{\beta}
\end{equation}

\noindent with $M_1$, $\epsilon_N$ and $\beta$ fitting functions that depend on redshift z. 
Using table 8 of Moster et al. (2018) we find to second order in (1+z)
\begin{eqnarray}
log (M_1/{\rm M}_{\odot}) & \approx & 11.62 + 0.1583 \times (1+z) - 0.01166 \times (1+z)^2 \nonumber \\
\epsilon_N & \approx & 0.09 + 0.06 \times (1+z) - 0.0048 \times (1+z)^2 \\
\beta & \approx & 2 - 0.24 \times (1+z) + 0.018 \times (1+z)^2 \nonumber
\end{eqnarray}
\noindent Combining equations 6 and 7 leads to
\begin{equation}
r_0 = 5 {\rm kpc} \left( \frac{M_*}{L_B} \right)_{\odot}^{-0.446} \left( 2 \epsilon_N \right)^{0.446}
\left( \frac{M_{vir}}{M_1} \right)^{0.446 \times \beta} \left( \frac{M_{vir}}{10^9 {\rm M}_{\odot}} \right)^{0.446}
\end{equation}

\noindent The red shaded area in the right panel of figure 1 shows the dependence of r$_0$ on M$_{vir}$
for redshifts z=0 (lower boundary) to z=8 (upper boundary), adopting (M$_*$/L$_B$)$_{\odot} = 1$. 
Core radii increase with dark halo virial mass following a power law: r$_0 \sim$ M$_{vir}^{\gamma}$.
The correlation is roughly linear with $\gamma$=1.0 for z=8, steepening somewhat to
$\gamma$=1.2 for low redshifts. 

Let us now compare these observations with the predictions for FDM cores as discussed in the introduction
(Chavanis 2011; Schive et al. 2014a,b; Schwabe et al 2016; Maleki et al. 2020). Rewriting equation 3 we get

\begin{equation}
\Sigma_{0} = 19 \left( \frac{10^{-22}{\rm eV}}{m} \right)^2 \left( \frac{{\rm kpc}}{r_0} \right)^{3} {\rm M}_{\odot} {\rm pc}^{-2}.
\end{equation}

\noindent In addition, equation 4 shows that  r$_0$ should decrease with halo virial mass as r$_0 \sim$ M$_{vir}^{-1/3}$.
The blue lines in figure 1 show these two FDM scaling relations for different dark matter masses m and z=0.
The thick line in each panel depicts the standard mass of m=$10^{-22}$eV. 
The predicted FDM scaling relations are in complete disagreement with the observations. For given m the core column density is predicted to 
steeply decrease with increasing r$_0$ while the observations show a constant $\Sigma_0$ (see also Deng et al. 2018).
In addition, FDM core radii r$_0$ should decrease with virial mass while the observations, 
combined with abundance matching, indicate a core radius that increases with M$_{vir}$.

One caveat in determining the correlation between core radius and virial mass is the assumption of a constant stellar mass-to-light ratio. 
For our purpose the absolute value of (M$_*$/L$_B$)$_{\odot}$ is not important as we are only interested in comparing trends with virial masses. However stellar mass-to-light
might also change systematically with M$_{vir}$. A detailed investigation is beyond the scope of this paper. However in order to evaluate this effect 
it is instructive to compare the predictions of equation 9
with the analyses of Salucci et al. (2007) who took a completely different approach. 
They determined dark halo virial masses from the kinematics of a large sample of disk galaxies,
adopting their universal rotation curve model and found r$_0 \approx$4.5$\times$(M$_{vir}/10^{11}$M$_{\odot}$)$^{0.58}$. The green line in the right panel 
of figure 1 shows this scaling relation. It is remarkable that both relationships are rather similar, given
the fact that the way how M$_{vir}$ is determined is very different.
The slope of the green line is somewhat less steep than the red area. It however confirms our conclusion that
the observationally inferred scaling relation between core radius and virial mass is opposite to the scaling relation expected for FDM cores.

\section{Introducing the redshift dependence of core properties}
So far we focussed on halos that continuously accumulate dark matter from the cosmic web till z=0.
As discussed earlier, satellites and cluster galaxies disconnect from the cosmic web at some z$>$0.
Their dark matter structure and by this also their soliton core properties are then likely to be frozen in, 
unless processes like tidal interactions with the surrounding
or gravitational interaction with the baryonic component within the galaxy change the halo core structure.
Here we neglect these secular processes and focus on the question whether introducing z (which marks the formation redshift of the halo)
as a second free parameter can bring soliton cores into agreement with the observed scaling relations.

Let us start with the core radius - virial mass relationship. As discussed in the previous section, the core radii of halos with
masses M$_{vir} \leq 5 \times 10^{11}$ M$_{\odot}$ should scale with virial masses as r$_0 \sim$ M$_{vir}^{\gamma}$ with 
$\gamma$ in the range of 0.6 to 1.2.
Soliton cores, on the other hand, are characterized by equation 4:
M$_{vir} \sim$ r$_0^{-3}\times$(1+z)$^{-1.5}$. Combining both relationships leads to

\begin{equation}
M_{vir} \sim (1+z)^{-\frac{1.5}{1+3\gamma}}
\end{equation}

\noindent Equation 11 provides interesting information about the formation redshift of
galaxies which is again strongly dependent on their environmental properties. For example, adopting $\gamma$ of order unity 
for our sample of galaxies, equation 11  predicts M$_{vir}\sim$(1+z)$^{-0.4}$.
The formation redshift dependence of the Salucci et al. (2007) sample with $\gamma$=0.6 is very similar with
M$_{vir}\sim$(1+z)$^{-0.5}$. In both cases, virial mass increases with decreasing z
which is consistent with cosmological hierarchical structure formation.
It would now be interesting to explore the effect of environment and by this
formation redshift on $\gamma$ in greater details and compare this with the predictions
of FDM. In summary, introducing
formation redshift, the second core-scaling relation can be brought into agreement with the FDM scenario.

The situation is however much more challenging for FDM when considering the first scaling relation of a constant core column density.
Equation 3 shows that the soliton core structure $\rho$(r) does not depend on z, M$_{vir}$ or any other halo property. 
It directly reflects the fundamental balance between the effects of the uncertainty principle and gravity. 
Through observations of halo cores one can therefore
determine the FDM particle mass m, independent of the core formation history or the
halo formation redshift. In principle one could also test the validity of the FDM model directly by comparing the observed density
distribution of dark matter cores with equation 3. Rewriting equation 10 leads to

\begin{equation}
\left( \frac{r_0}{{\rm kpc}} \right)^3 = 0.25 \left( \frac{75 {\rm M}_{\odot} {\rm pc}^{-2}}{\Sigma_0} \right) \left( \frac{10^{-22}{\rm eV}}{m} \right)^2
\end{equation}

\noindent Cores with the observed constant column density therefore should all have the same radii, r$_0$.
As shown by figure 1 this is in clear contradiction with the observations where the core radii for given $\Sigma_0$ 
change by at least two orders of magnitude.
The only possible solution would then be that FDM consists of a population of particles with different masses m. 
In addition, it would require that these particles are not
distributed randomly but that galaxies with larger core radii are populated by FDM particles with smaller average masses 
such that $\langle$m$\rangle\sim$r$_0^{-3/2}$. One cannot rule out this scenario. It however appears highly constructed and therefore unrealistic.

In summary, even if we include the redshift dependence of core formation as an additional free parameter, soliton
cores are not able to explain both observed core scaling relations simultaneously.

\noindent 

\section{Conclusions and Discussion}

Observed dark matter core scaling relations provide powerful constraints for models of dark halo formation and the nature
of the dark matter particle. Here we focussed on two relationships
that have to be fulfilled simultaneously by any theoretical model of core formation: a constant core column
density and a core radius that increases with virial mass.
Applied to FDM we have shown that soliton cores with a structure and redshift dependence as found by Schive et al. (2014a,b) 
can in principle explain the observed r$_0$-M$_{vir}$ relationship. In fact, within the FDM scenario
this relationship reflects the dependence of the halo virial masses on their formation redshift
which might change with galaxy type and cosmic environment.
FDM however cannot explain the origin of the observed universal core column density.
According to equation 12, cores with similar column densities should all have the same radii, r$_0$. 
This is in contradiction with the observations that the core radii for given $\Sigma_0$ change by at least two orders of magnitude (figure 1).
FDM therefore appears to be ruled out as an explanation for the origin of the observed dark matter cores. 

One should note, that this conclusion relies
on the results of the cosmological simulations by Schive et al. (2014a,b). It would now be
interesting to refine their predicted core structure and its redshift dependence with a larger
sample of high-resolution simulations.
One should also explore the gravitational interaction of FDM cores with baryons and here focus especially 
on violent relaxation effects from strong gravitational perturbations
that could pump energy into FDM cores, reducing their degeneracy and compactness
and increasing their radii. Finally, 
outside a soliton core, FDM behaves like ordinary dark matter and one expects a transition
to the standard NFW profile. The question then arises whether the observed cores actually trace
this transition region, rather than the soliton core itself.

The observational evidence for dark matter cores in halos ranging from dwarf galaxies to galaxy clusters 
is a very active research field and still a matter of controversal debate
(del Popolo \& Le Delliou 2017). For example, Zhou et al (2020) find
a weak dependence of core surface density on galaxy luminosity L: $\Sigma_0 \sim L^{0.13}$.
Chan (2014) presents a systematic study of galaxy clusters and finds cores with surface densities that scale with core radius as
$\Sigma_0 \sim r_0^{-0.46}$.
Gopika \& Desai (2020) lateron analyse 12 relaxed Chandra X-ray clusters and find that to good approximation
the surface density is constant with $\Sigma_0 \sim r_0^{-0.08}$. At the low-mass end, Hayashi et al. (2020) analyse eight classical
dwarf spheroidal galaxies and find a diversity of core structures with many actually favoring cuspy profiles. In their analyses, Hayashi
et al. (2020) assume dynamical equilibrium. Satellite galaxies are however known to be strongly tidally perturbed.
It would now be important to investigate the effects of tides on FDM halos in greater details.
If the cuspy central structure of some dwarf spheroidals cannot be reproduced it
would generate even more problems for FDM, raising the question how cusps could be generated 
if flat soliton cores are a fundamental property of every FDM halo. In summary however, despite the on-going discussion about
the frequency of cores and their scaling relations, all observations
agree that whenever cores are detected their radii can vary by orders of magnitude for given core surface density. This
is in contrast to the predictions of the FDM scenario.
No observation has yet found the steep dependence of $\Sigma_0 \sim$ r$_0^{-3}$ (equation 12) which is a fundamental property of FDM cores.

If FDM turns out not to explain the origin of the observed dark halo cores it would of course not rule out the
existence of the FDM particle in general. 
As shown by figure 1, a constant core column density $\Sigma_0$ has been
confirmed down to core radii of order 300 pc, corresponding to dark halo virial masses of order 5$\times10^9$-1$\times10^{10}$M$_{\odot}$.
The core scaling relations could in principle change below this observational limit and follow for example 
the FDM prediction $\Sigma_0\sim$r$^{-3}$. This would require axion masses to be larger than 
m$_{min} \approx 3 \times 10^{-22}$eV. In this case the soliton cores would be too small
to be detectable, at least with current observational techniques. The observed constant density cores 
with radii ranging from 300 pc to more than 10 kpc must then be the result of other processes.

Such a scenario is possible, in principle. The question however arises whether one should focus on FDM as a dark matter candidate
if its major motivation, namely to explain the origin of the observed dark matter cores, is not valid anymore.

\acknowledgments
I thank the Harvard Center for Astrophysics, Cambridge for their invitation and for inspiring discussions that started this project.
Special thanks to Philip Mocz for clarification of the FDM core redshift dependence and to the referee for constructive suggestions that
made the paper more clear.
Thanks also to Avi Loeb, Wayne Hu, Mohammad Safarzadeh, Jens Niemeyer, Paolo Salucci, Volker Springel and Antonino del Popolo
for helpful comments and to John Kormendy for providing his compilation of dark halo core properties.
This research was supported by the Excellence Cluster ORIGINS which is funded by the Deutsche Forschungsgemeinschaft 
(DFG, German Research Foundation) under Germany\textsc{\char13}s Excellence Strategy - EXC-2094 - 390783311.

\end{document}